\documentclass[%
 amsmath,amssymb,
 aps,
]{revtex4-2}

\usepackage{graphicx}
\usepackage{dcolumn}
\usepackage{bm}

\usepackage{multirow}%
\usepackage{amsmath,amssymb,amsfonts,bm}%
\usepackage{amsthm}%
\usepackage{mathrsfs}%
\usepackage[title]{appendix}%
\usepackage{xcolor}%
\usepackage{textcomp}%
\usepackage{booktabs}%
\usepackage{algorithm}%
\usepackage{algorithmicx}%
\usepackage{algpseudocode}%
\usepackage{listings}%
\usepackage{braket}
\usepackage{comment}
\usepackage[normalem]{ulem}
\newcommand{\YN}[1]{{\color{red} #1}}

\begin{document}

\preprint{APS/123-QED}

\title{Variational readout through quantum teleportation}

\author{Yohei Nishino$^{1,2}$}
 \email{yohei.nishino@grad.nao.ac.jp}

\affiliation{$^1$Department of Astronomy, University of Tokyo, Bunkyo, Tokyo 113-0033, Japan}
\affiliation{$^2$Gravitational Wave Science Project, National Astronomical Observatory of Japan (NAOJ), Mitaka City, Tokyo
181-8588, Japan}

\date{\today}

\begin{abstract}
Sensitivity of gravitational-wave detectors (GWDs) is constrained at low frequencies by quantum radiation-pressure noise, a manifestation of the measurement's back action. One strategy to mitigate this back action involves employing variational readout, which entails cross-correlating the measurement shot noise with radiation-pressure noise. Prior research has demonstrated that variational readout in GWDs can be accomplished through the use of a filter cavity. However, current gravitational-wave detectors necessitate filter cavity lengths on the order of $\sim 100$ meters, with future detectors anticipated to reach lengths of a few kilometers. This paper introduces a novel approach to variational readout utilizing principle of quantum teleportation, which eliminates the need for a filter cavity.
\YN{This document will not be published any journals, since it turned out to be not very useful for the real detectors. The main reason is that one does not need the EPR entanglement to obtain the same result, which does not even shows true variational readout. In other words, both schemes do not exceed the EPR squeezing. Nevertheless this document will be uploaded to stimulate discussions how to realize "true" back action evasion, such as variational readout or speed meter, via quantum entanglement.}
\end{abstract}

\maketitle


\section{\label{sec:level1} Introduction}
Gravitational wave detectors face a fundamental limitation due to quantum noise, encompassing shot noise and quantum radiation pressure noise (QRPN). While shot noise represents the intrinsic uncertainty of measurement, QRPN is back action of measurements, creating a trade-off called the standard quantum limit (SQL).

Numerous strategies have been proposed to overcome this limit, including Quantum Non-Demolition (QND) measurements. In a seminal work~\cite{PhysRevD.65.022002} (called KLMTV hereafter), the injection of a squeezed state of light was suggested. With a fixed angle injection, given one intends to reduce shot noise, QRPN is increased. To address this, they proposed frequency-dependent squeezing (FDS) with filter cavities, a method already implemented in detectors such as LIGO and Virgo, and planned for future use in Einstein Telescope and Cosmic Explorer.

Additionally, KLMTV introduces variational readout as another QRPN reduction method, adopting a post-filtering approach (originally proposed in~\cite{VYATCHANIN1995269}). This technique exploits the correlation between shot noise and QRPN, canceling out the latter. The variational readout interferometer, in its ideal form, is back-action-free, allowing sensitivity improvement by increasing laser power or squeezing levels. Back-action-evading measurements have broader implications and are studies not only in gravitational-wave detection~\cite{VYATCHANIN1993772218, VYATCHANIN1994492, VYATCHANIN1996834690, VYATCHANIN19968261007, DANILISHIN2002547, DANILISHIN2000123}, but also quantum mechanics in the macroscopic world~\cite{Murch2008, PhysRevX.7.021008, PhysRevX.7.031055, PhysRevLett.117.140401, andp.201400099, M_ller_2017}. 

The conventional scheme in KLMTV involves an additional long-baseline  and low-loss Fabry-P\'erot cavity to narrow its bandwidth, presenting challenges of costs. Our paper proposes a new approach to variational readout using quantum teleportation principles. A previous studies~\cite{nishino2024frequencydependent} explored the collaboration between gravitational-wave detectors and quantum teleportation, repurposing the interferometer as pre-filter cavities. Extending this idea, we introduce repurposing the interferometer as a post-filter cavity, offering the potential for back-action-free measurements in gravitational-wave detectors without the need for filter cavities, promising enhanced sensitivity and cost-effective advancements.

\section{\label{sec:level2} A brief summary}

\begin{figure*}
    \centering
    \includegraphics[scale=0.9]{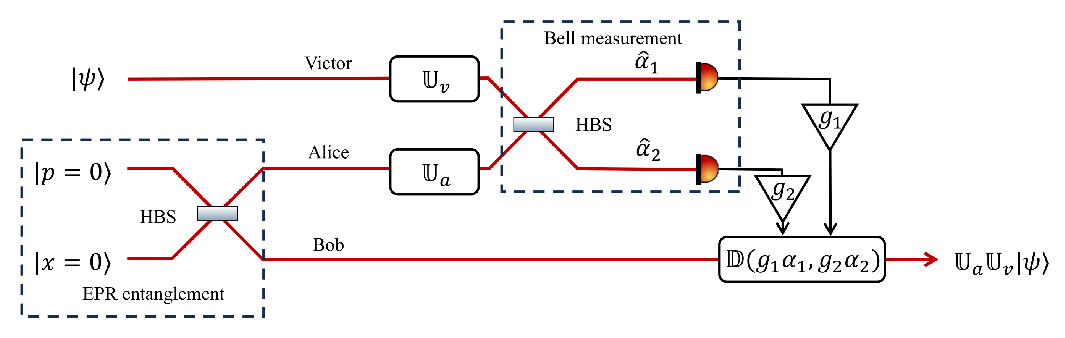}
    \caption{Flowchart of continuous variable teleportation. When $\mathbb{U}_v$ is ponderomotive squeezing and $\mathbb{U}_a$ is a quadrature rotation, the output state is equivalent to the variational readout scheme.}
    \label{fig:Fig_CV_teleportation_QTVR}
\end{figure*}

In a previous study~\cite{nishino2024frequencydependent}, frequency-dependent pre-filtering is achieved by transforming the quantum states of Victor, Alice, and Bob during the teleportation process.
By altering the order of these transformations, post-filtering, also known as "variational readout," can be attained. As illustrated in the flowchart in Fig.~\ref{fig:Fig_CV_teleportation_QTVR}, Victor undergoes ponderomotive squeezing, and Alice experiences a quadrature rotation. In tuned interferometers, Bob is detected without being injected into the interferometer. After post-processing with Wiener filters, the teleported state is equivalent to variational readout.

\begin{figure}[b]
    \centering
    \includegraphics[scale=0.5]{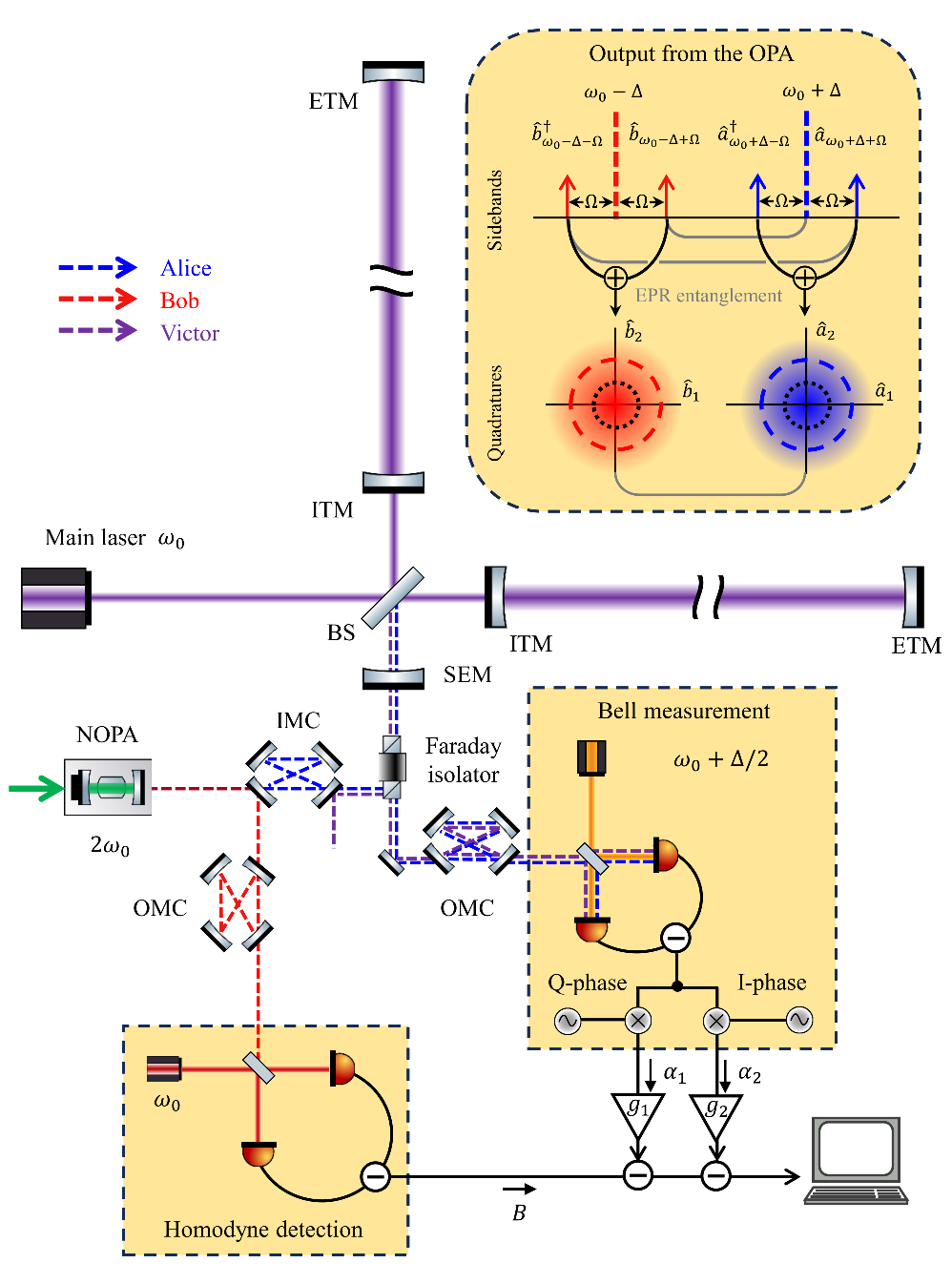}
    \caption{Configuration of quantum-teleportation variational readout.}
    \label{fig:Fig_configuration_QTVR}
\end{figure}
Fig.~\ref{fig:Fig_configuration_QTVR} depicts the actual implementation in gravitational-wave detectors. The optical parametric amplifier (OPA) is pumped at a frequency of $2\omega_0$, twice the frequency of the main laser. The two-mode EPR entanglement-Alice and Bob—are generated through parametric down-conversion in a non-linear crystal. The central frequencies of Alice and Bob are $\omega_0\pm\Delta$, respectively, from the law of energy conservation.

Bob is reflected by the Input-Mode Cleaner (IMC) and measured by a homodyne detector. Victor and Alice are injected into the interferometer through the Faraday isolator. The output is filtered by the Output-Mode Cleaner (OMC) and detected through Bell measurement. In Bell measurement, the Bell observable vector $\pmb{\hat{\alpha}}$ is defined as:
\begin{align}
    \pmb{\hat{\alpha}} = \begin{pmatrix}
       \hat{\alpha}_1\\
       \hat{\alpha}_2
    \end{pmatrix}
    = \frac{1}{\sqrt{2}} \begin{pmatrix}
        \hat{V}_1-\hat{A}_1 \\
        \hat{V}_2+\hat{A}_2
    \end{pmatrix}.
\end{align}
This can be achieved by a local oscillator at the frequency of $\omega_0+\Delta/2$, which is the central frequency of Victor and Alice. The three measurement results are combined with Wiener filters, ultimately achieving back-action evasion.

\section{\label{sec:level3} Analysis of variational readout}
\subsection{Different functionality of interferometer for Victor and Alice}
Victor perceives the interferometer as an active cavity, manifesting the ponderomotive-squeezing effect characterized by the Kimble factor $\mathcal{K} = \frac{2\Theta \gamma}{\Omega^2(\Omega^2+\gamma^2)}$ with $\Theta = \frac{8\omega_p I_c}{McL}$. The input and output relation of Victor is expressed as follows~\cite{PhysRevD.65.022002}:
\begin{align}
    \pmb{\hat{V}} = \mathbb{U}_v \pmb{\hat{v}} = e^{2i\beta_b}\begin{pmatrix} 
        1 & 0 \\
        -\mathcal{K} & 1 
    \end{pmatrix}
    \begin{pmatrix}
        \hat{v}_1 \\
        \hat{v}_2
    \end{pmatrix}
    + \sqrt{\mathcal{K}}e^{i\beta_v} \begin{pmatrix}
        0\\
        1
    \end{pmatrix} \frac{h}{h_\mathrm{SQL}}, \label{eq:V}
\end{align}
where $\beta_v$ is a common quadrature rotation, and $h_\mathrm{SQL}=\sqrt{\frac{8\hbar}{ML^2\Omega^2}}$ is the standard quantum limit for a free test mass.

The interferometer functions as a passive cavity for Alice, i.e.:
\begin{align}
    \pmb{A} = \mathbb{U}_a \pmb{\hat{a}} = e^{2i\beta_a}\begin{pmatrix}
        \cos\theta_a & -\sin\theta_a \\
        \sin \theta_a & \cos\theta_a
    \end{pmatrix}
    \begin{pmatrix}
        \hat{a}_1 \\
        \hat{a}_2
    \end{pmatrix}.
\end{align}
Bob is detected without entering from the AS port, resulting in the output being identical to the input:
\begin{align}
    \pmb{\hat{B}} = \pmb{\hat{b}}.
\end{align}
In the actual configuration, the phase quadrature is measured, meaning $\hat{B} = \hat{b}_2$.

\subsection{Post processing}
\begin{figure}
    \centering
    \includegraphics[scale=0.28]{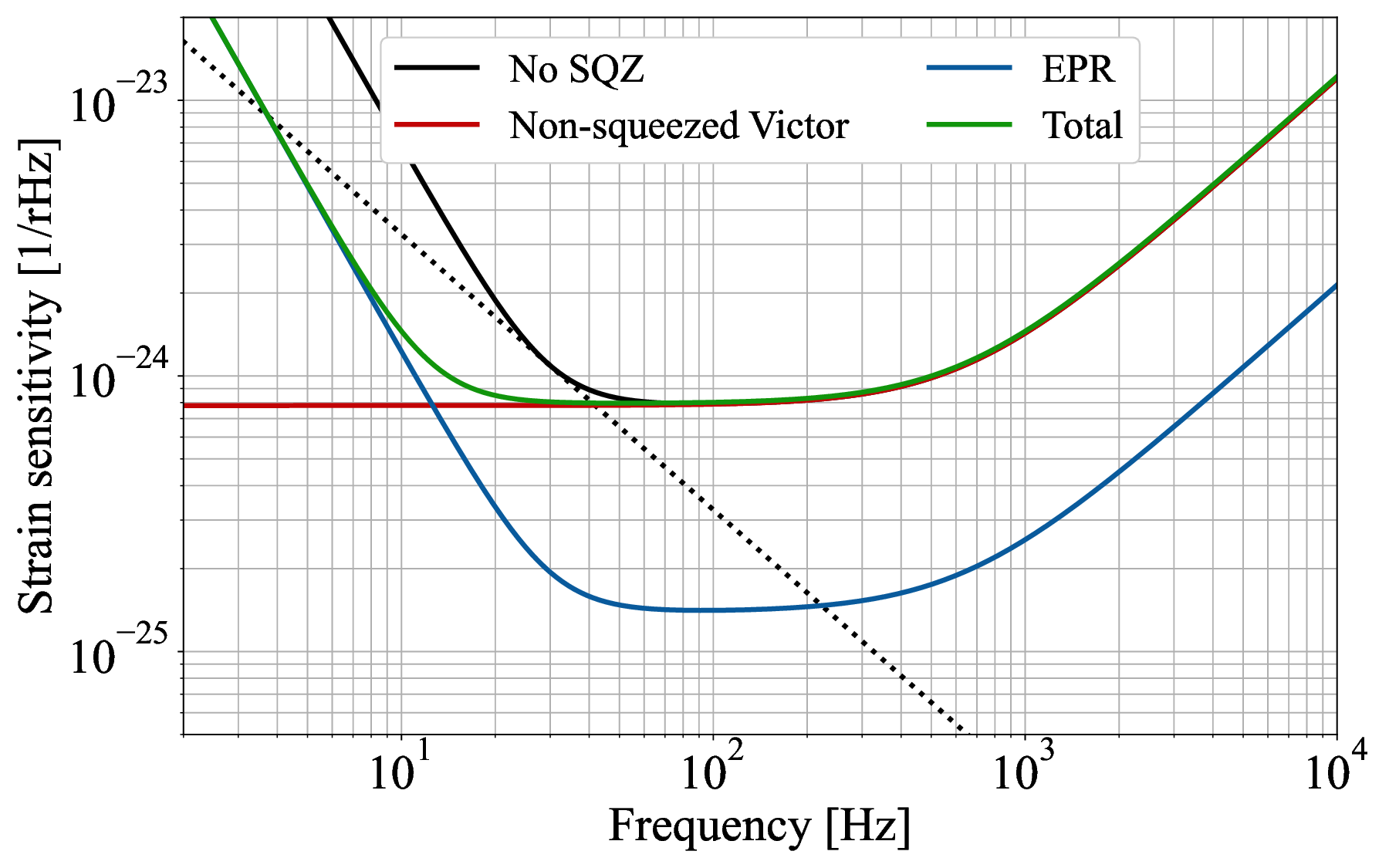}
    \caption{Comparison of noise components in Eq.~(\ref{eq:S_h}) when Victor is not squeezed}
    \label{fig:ss_vab}
\end{figure}

The Bell measurement outputs $\hat{\alpha}{1,2}$ are subtracted from Bob's field with filters $g{1,2}$ as follows:
\begin{align}
    \hat{B}^g = \hat{B} - g_1\hat{\alpha}_1 - g_2\hat{\alpha}_2.
\end{align}
Since the gravitational-wave signal is imprinted in the phase quadrature of Victor's field, $V_2$ in Eq.~\ref{eq:V}, and is also filtered out by $g_2$, the final strain sensitivity is calculated as:
\begin{align}
    S_h = \frac{h_\mathrm{SQL}^2}{2\mathcal{K}}\frac{S_{\hat{B}^g\hat{B}^g}}{|g_2|^2}, \label{eq:S}
\end{align}
where 
\begin{align}
    S_{\hat{B}^g\hat{B}^g} &= S_{\hat{B}\hat{B}} + |g_1|^2 S_{\hat{\alpha}_1\hat{\alpha}_1} + |g_2|^2 S_{\hat{\alpha}_2\hat{\alpha}_2} \notag \\
    &-g_1^*S_{\hat{B}\hat{\alpha}_1}-g_1S_{\hat{\alpha}_1\hat{B}} -g_2^*S_{\hat{B}\hat{\alpha}_2}-g_2S_{\hat{\alpha}_2\hat{B}} \notag \\
    & + g_1g_2^*S_{\hat{\alpha}_1\hat{\alpha}_2} + g_1^*g_2S_{\hat{\alpha}_2\hat{\alpha}_1}. \label{eq.Spectrum}
\end{align}
Eq.~(\ref{eq:S}) reaches its minimum when $g_{1,2}$ are Wiener filters, defined as:
\begin{align}
    g_1 &= \frac{S_{\hat{B}\hat{\alpha}_1}S_{\hat{\alpha}_2\hat{B}}-S_{\hat{\alpha}_2}\hat{\alpha_1}S_{\hat{B}\hat{B}}}{S_{\hat{\alpha}_2\hat{B}S_{\hat{\alpha}_2\hat{\alpha}_2}}-S_{\hat{\alpha}_1\hat{B}S_{\hat{\alpha}_2\hat{\alpha}_1}}} \label{eq:g1} \\
    g_2 &= \frac{S_{\hat{\alpha}_1\hat{\alpha}_1}S_{\hat{B}\hat{B}}-|S_{\hat{\alpha}_1\hat{B}}|^2}{S_{\hat{\alpha}_2\hat{B}S_{\hat{\alpha}_2\hat{\alpha}_2}}-S_{\hat{\alpha}_1\hat{B}S_{\hat{\alpha}_2\hat{\alpha}_1}}} \label{eq:g2} 
\end{align}
The power (cross) spectral densities of the outputs are calculated as:
\begin{align*}
    S_{\hat{B}\hat{B}} &= \cosh2r,\\
    \ S_{\hat{\alpha}_1\hat{\alpha}_1} &= \frac{1}{2}(S_{\hat{v}_1\hat{v}_1}+\cosh{2r}), \\
    S_{\hat{\alpha}_2\hat{\alpha}_2} &= \frac{1}{2}(\mathcal{K}^2S_{\hat{v}_1\hat{v}_1}+S_{\hat{v}_2\hat{v}_2}+\cosh2r),\\ S_{\hat{\alpha}_1\hat{\alpha}_2} &= -\frac{e^{1}}{2}\mathcal{K}S_{\hat{v}_1\hat{v}_1} \\
    S_{\hat{B}\hat{\alpha}_1} &= -\frac{1}{\sqrt{2}}e^{2i(\beta_b-\beta_a)}\sin\theta_a\sinh2r \\
    S_{\hat{B}\hat{\alpha}_2} &= -\frac{1}{\sqrt{2}}e^{2i(\beta_b-\beta_a)}\cos\theta_a\sinh2r.
\end{align*}
Substituting them into Eqs.~(\ref{eq:g1}, \ref{eq:g2}) and tuning Alice's rotation $\theta_a=\mathrm{arctan}\mathcal{K}$, Eq.~(\ref{eq:S_h}) leads:
\begin{align}
    S_h^g = \frac{h_\mathrm{SQL}^2}{2\mathcal{K}}\left(S_{\hat{v}_2\hat{v}_2} + \frac{1+\mathcal{K}^2}{\cosh{2r}}\right). \label{eq:S_h}
\end{align}
The first term represents the teleported noise of Victor, exhibiting only its phase quadrature in detection. It has the same structure as the variational readout~\cite{PhysRevD.65.022002}. As shown in the red curve in Fig.~\ref{fig:ss_vab}, back-action noise is totally canceled out at low-frequency region, even below the SQL. This corresponds when Victor is a normal vacuum entering from the AS port in Fig.~\ref{fig:Fig_configuration_QTVR}.

\begin{figure}
    \centering
    \includegraphics[scale=0.28]{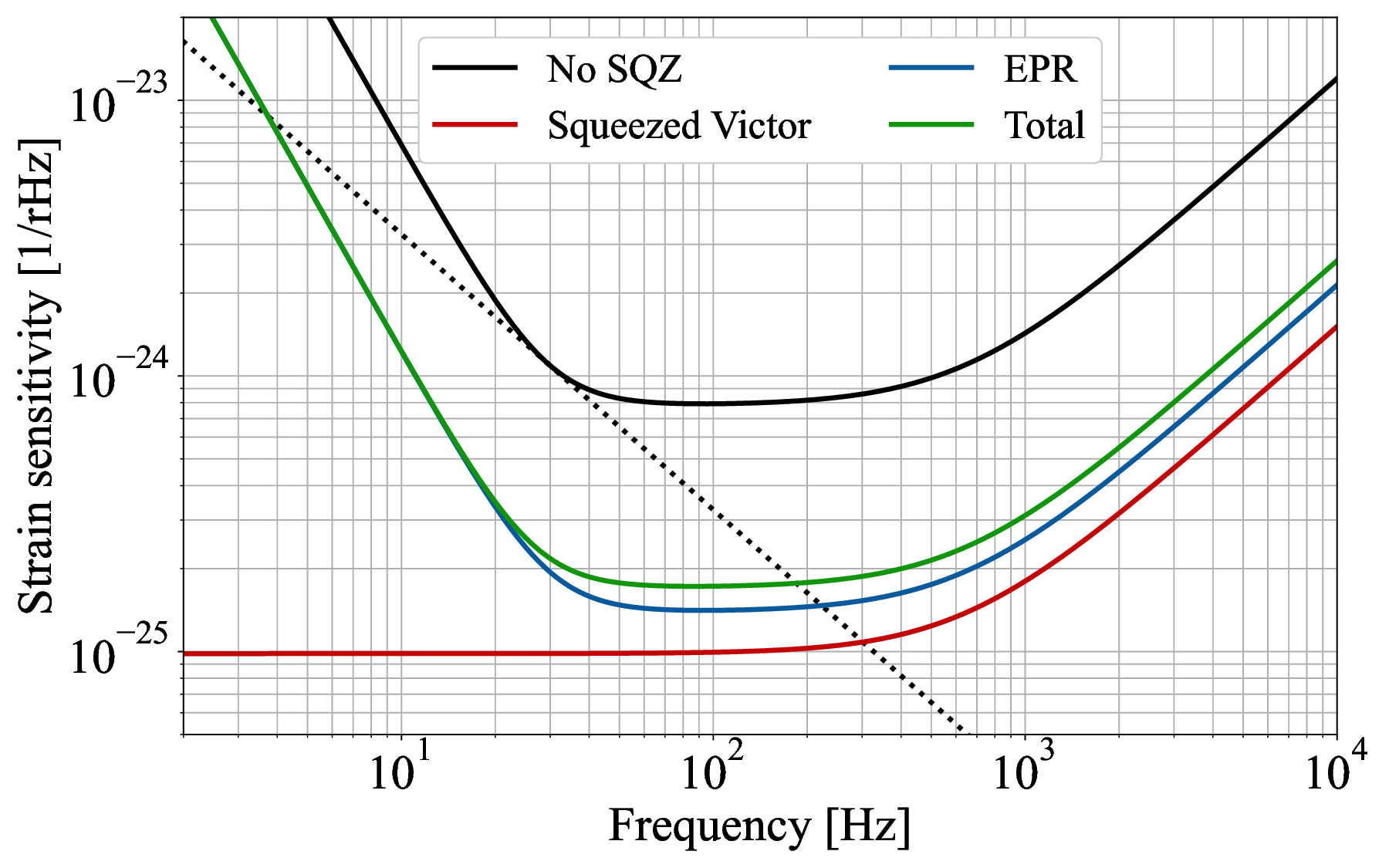}
    \caption{Comparison of noise components in Eq.~(\ref{eq:S_h}) when Victor is squeezed}
    \label{fig:ss_svab}
\end{figure}

The influence of the finite squeezing level appears in the second term, introducing back-action noise (refer to the blue curve in Fig.~\ref{fig:ss_vab} and~\ref{fig:ss_svab}). This contribution originates from Alice and Bob, exhibiting the same profile as conventional quantum noise multiplied by a factor of $1/\cosh{2r}$ and restricting sensitivity at very-low frequencies.

It is also possible to manipulate the phase quadrature of Victor through squeezing. The squeezed state is generated at half the frequency of the pumping laser to the OPA. However, when Victor's squeezing level aligns with the entanglement, this second term becomes dominant at low frequencies. Remarkably, this noise level aligns precisely with EPR squeezing~\cite{Ma_2017}, as depicted in the blue curve in Fig.~\ref{fig:ss_svab}.

\begin{table}[]
\caption{Parameters for ETHF-like detectors in three configurations}\label{tab1}%
\begin{tabular}{p{4.cm}|p{2.cm}|p{2.cm}}
    Parameters & Baseline FDS & EPR/QTVR  \\ \hline\hline
    Filter cavity length & 1 km & - \\ \hline
    Arm round trip loss & \multicolumn{2}{c}{80 ppm} \\ \hline
    SEC loss & \multicolumn{2}{c}{1000 ppm} \\ \hline
    Injection loss & \multicolumn{2}{c}{3 \%} \\ \hline
    Readout loss & \multicolumn{2}{c}{3 \%} \\ \hline
    FC round-trip loss & 45 ppm &  - \\ \hline
    Squeezer noise RMS\footnotemark[1] & \multicolumn{2}{c}{10 mrad} \\ \hline
    Local oscillator RMS & \multicolumn{2}{c}{10 mrad} \\ \hline
    SEC length RMS & \multicolumn{2}{c}{1 pm} \\ \hline
    Filter cavity length RMS & 1 pm & - \\ \hline
    Detuning $\Delta$  & - & $\sim 49.4$~MHz \\ \hline
    Squeezing level & \multicolumn{2}{c}{-18 dB} \\ \hline
\end{tabular}
\footnotetext[1]{Root mean square}
\end{table}

\section{Comparison with EPR squeezing}
\begin{figure}[h]
    \centering
    \includegraphics[scale=0.28]{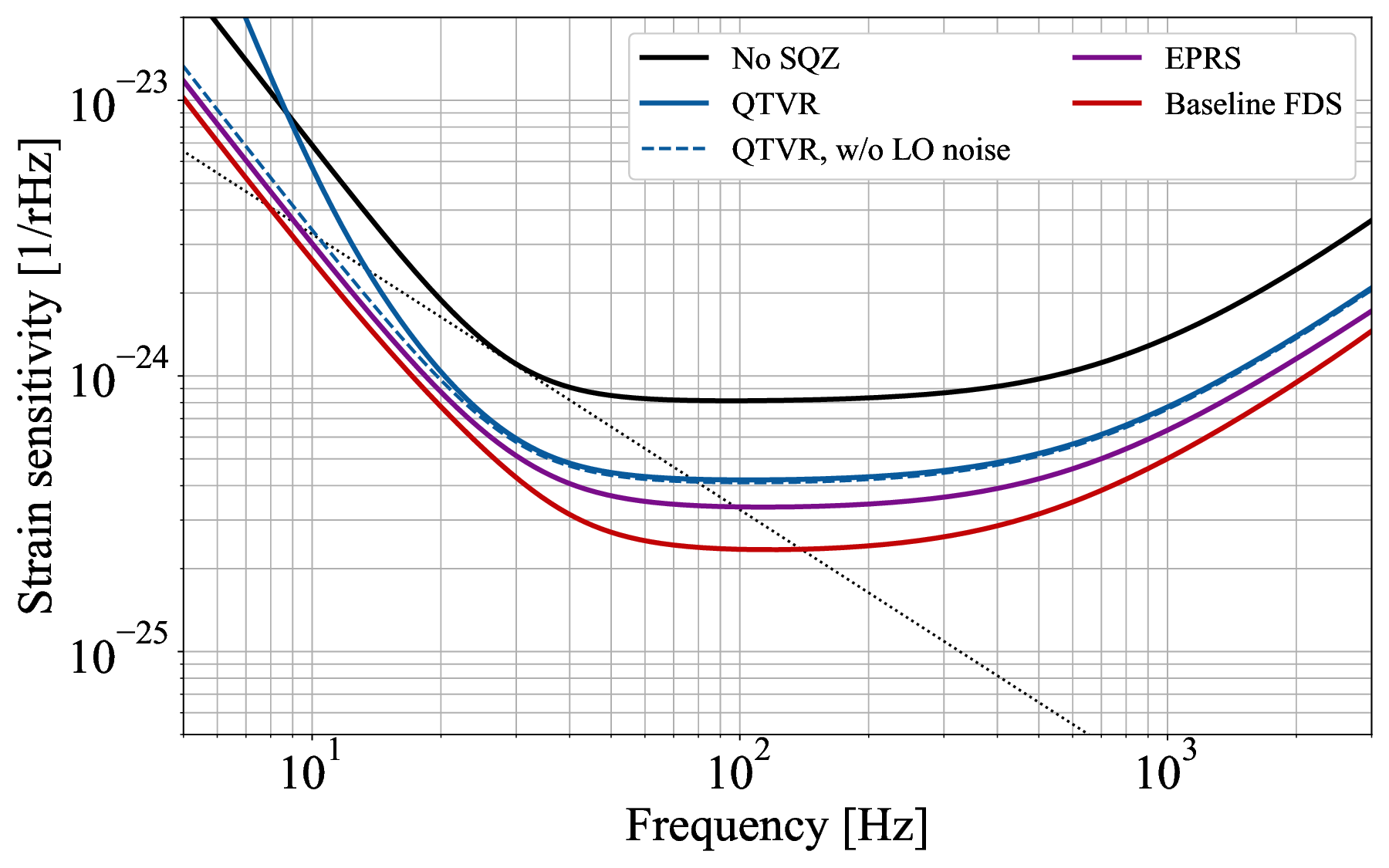}
    \caption{Comparison of QTVR with the baseline FDS and EPRS. The dashed curve is QTVR without the local-oscillator noise.}
    \label{fig:ss_EPR_comparison}
\end{figure}

In Fig.~\ref{fig:ss_EPR_comparison}, we compare the performance of three configurations: the baseline frequency-dependent squeezing (FDS), EPR squeezing (EPRS), QT variational readout (QTVR). We used parameters of the high-frequency detector of the Einstein Telescope xylophone detector. We take into account several loss sources: (1) 3 \% injection loss including the OPA cavity loss and Faraday isolator loss. (2) 80 ppm round trip loss in the arm cavities. (3) 1000 ppm loss in the signal-extraction cavity. (4) 3 \% readout loss including the photo detector, Faraday, and output-mode cleaner losses. Those losses introduce additional vacuum which has no correlation with the squeezing vacuum. Also, we considered three sources of phase noise in root-mean square (RMS) values: (1) 10 milliradians of squeezer noise relative to the main laser. (2) 10 mrad of local oscillator noise. (3) 1 pm of the SEC cavity length noise (see Table~\ref{tab1}). For the baseline FDS, an 80 ppm round-trip loss and 1 pm length noise of the filter cavity are taken into account.

Two important points should be noted about Fig.~\ref{fig:ss_EPR_comparison}: firstly, the sensitivity of QTVR is inferior to EPRS across the whole frequency range. This is because the second term of Eq.~(\ref{eq:S_h}) has the same level of uncertainty as EPRS. Therefore with the same amount of squeezing level for Victor and EPR sideband, the strain sensitivity of EPRS $S_{h,\mathrm{EPRS}}^g$ serves a lower bound for that of QTVR $S_{h,\mathrm{QTVR}}^g$, i.e., $S_{h,\mathrm{EPRS}}^g < S_{h,\mathrm{QTVR}}^g$. 

Secondly, QTVR steep a steep noise below $\sim 20$ Hz. This is because the correlation between shot noise and radiation pressure noise is highly susceptible to dephasing caused by the local-oscillator noise (compare the blue-solid and dashed curve).

\section{conclusion}
This investigation demonstrates the feasibility of variational readout without the need for an additional filter cavity by leveraging the principles of quantum teleportation. As depicted in Fig.~\ref{fig:ss_vab}, quantum radiation pressure noise is mitigated, enabling a surpassing of the Standard Quantum Limit (SQL). However, at very low frequencies, the impact of entanglement imperfections becomes tangible. 

One can utilize a squeezed state of light generated from the same OPA as the entanglement. In this case, the noise is reduced in broadband by tuning the fixed squeezing angle. However, at any frequency, the sensitivity cannot be better than EPR squeezing. This is because the fundamental noise limit of this scheme arising from the imperfection of the EPR entanglement, have exactly the same shape as the EPR squeezing. In addition to that, our scheme has the noise of Victor, which does not allow to surpass the noise level of EPR squeezing.

\bibliography{references}
\end{document}